\documentclass[conference]{IEEEtran}
\usepackage{graphicx}
\usepackage{multirow} 
\begin{document}
\title{\textbf{\Large{Packet-pair technique for available bandwidth estimation in IPv6 network}}}

\author{\IEEEauthorblockN{T.G. Sultanov}
\IEEEauthorblockA{\textit{Samara State Aerospace University}\\
\textit{Moskovskoe sh., 34,}\\
\textit{Samara, 443086, Russia}\\
\textit{E-mail: tursul@rambler.ru}}
\and
\IEEEauthorblockN{P.V. Veselovskiy}
\IEEEauthorblockA{\textit{Samara State Aerospace University}\\
\textit{Moskovskoe sh., 34,}\\
\textit{Samara, 443086, Russia}\\
\textit{E-mail: pavelv.ssau@yandex.ru}}}

\maketitle

\begin{abstract}
%\boldmath
This paper presents experimental checking of the model for measuring available bandwidth in IPv6. The experiment was performed using a measuring infrastructure RIPE test box, ensuring precision accuracy. The experimental results showed that to increase the accuracy of available bandwidth, we need to neutralize the effect of the variable part of the delay by increasing the number of measurements. Finally, we made the computer simulation, which allowed us to establish a dependence between the measurement error of the available bandwidth and the number of measurements.
\\
\end{abstract}


\IEEEpeerreviewmaketitle

\section{Introduction}
IP version 6 (IPv6) \cite{s12}, is a new version of Internet Protocol, designed as a successor to IP version 4 (IPv4) \cite{s13}. First of all, transition from IPv4 to IPv6 will result in enlarging address space capabilities, simplification of format of addressing header, reduction of number of routing tables, also improvement of security provision, effective mechanisms for quality of service (QoS) and mobile devices support. Therefore, a priority for further development of the Internet is introducing a new protocol version IPv6, whose basic specifications have been developed \cite{s10,s12,s16}. Note, that at the moment it is getting harder and harder to get IPv4 addresses due to exhaustion of available RIPE address space and the regional Internet registries \cite{s8}. The last part of IANA IPv4 addresses has been allocated, while this paper was in writing. Despite the fact that the IPv6 is based on an earlier version of the protocol, IPv6 should be seen as a new protocol that requires active investigation.

In Russia, the State Institute of Information Technologies and Telecommunications Computer Science, Institute of Organic Chemistry, Yaroslavl State University and the Russian Institute for Public Networks have already joined their efforts to create a network segment, which is based on IPv6 within single-space networks RUNNet, FREEnet and RBNet.

However, the introduction of IPv6 is slower than expected, and transition mechanisms play an important role, among which the address translation DNS64 and NAT64 seem to be perspective technology. This technology allows the use of the services of an older version of protocol in the global network from the local network, configured in accordance with the protocol of IPv6. This technology has already been working; for example, the project Ecdysis (NAT64 implementation of open source) has shown that you can make your network operate in IPv4, and in IPv6 \cite{s5}.

To have a correct idea of the Russian segment of the global network, we must have a connectivity map, and such work is already underway. In 2010, Samara State Aerospace University (SSAU) became a member of the program RIPE Atlas \cite{s14}, which will collect statistics about connections from mobile measuring devices in IPv4, and in IPv6 from the whole world. Now there are two connected and working measuring devices: in the SSAU and in Togliatti State University (TSU). The map, placed on the RIPE Atlas site, represents information about all active \textit{probes}.

Block (/48) IPv6 addresses are reserved in SSAU, and work is underway to transition to the new version of the protocol with all of the above innovative technologies. With the introduction of a new version of the protocol, experiments were made to measure quantities that characterize the performance of an IPv6 network. One of these important metrics is the \textit{available bandwidth} \cite{s11}.

To carry out precision experiments to evaluate network performance, as well as to develop new additions to standards, we should use a modern measuring infrastructure. In SSAU for several years the measurement system RIPE test box was rolled out under the grant RFBR 06-07-89074 \cite{s1}. It collects data, describing the main network parameters such as delay with a microsecond accuracy, variation of delay (\textit{jitter}), the routing path (\textit{trace route}), etc. These data enable us to study the relation between available bandwidth and the basic network parameters.

In paper \cite{s17} a new model of measuring the available bandwidth and the results of experiments in an IPv4 network were presented. In this connection it seems interesting to perform measurements in the networks of IPv6, using the measurement infrastructure RIPE test box. In the section "The model and scheme of experiment" a description is given of such an experiment, in the section "Processing of experimental results" - the results of measurements.

\section{The model and scheme of experiment}
In this section we would like to give a brief description of the model of measuring available bandwidth and the scheme of an experiment for checking this model.

\begin{figure*}
\centering
\includegraphics[height=6cm]{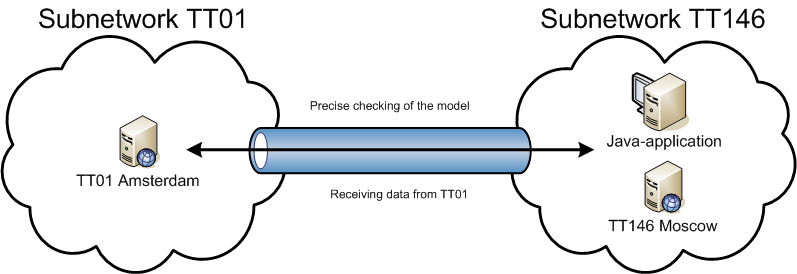}
\caption{Scheme of the experiment}
\label{f1}
\end{figure*}

In paper \cite{s17} it was suggested to test the network link with packets of different sizes \cite{s1}, so that the packet size varies up to the maximum possible value without possible router fragmentation. This method allows us to find a way to eliminate the measurement limitations of the variable delay component, $d_{var}$. The equation for measuring the available bandwidth could be modified to a suitable form:

 	\begin{equation}
  B_{av}=\frac{W_2-W_1}{D_2-D_1}
  \label{e1}
	\end{equation}
where $B_{av}$ is the available bandwidth, $W_1$,$W_2$ is the sizes of the transmitted packet and $D_1$,$D_2$ is the network packet delays (one-way delay).

As part of the experiment we made precise testing of a large number of packets of 100 and 1100 bytes, using a well-known system of active measurements, a RIPE test box. The number of measurement systems in the global measurement infrastructure reaches 80; these points cover the major world centers of the Internet, reaching their highest density in Europe. The measurement error of the packet delay is 2--12 microseconds \cite{s6}. In order to prepare the experiments, three test boxes were installed in Moscow, Samara and Rostov on Don in Russia from 2006 to 2008, supported by RFBR grant 06-07-89074. Measuring points in Moscow and Amsterdam were already working in IPv6, therefore, for further analysis we collected several data sets in both directions Moscow -- Amsterdam (tt146.ripe.net -- tt01.ripe.net), containing up to 2000 measurements. Data from measuring points were collected by using protocol \textit{telnet} to port 9142. Server tt01.ripe.net has public access and access to data from tt146.ripe.net is possible from the network segment FREEnet, Moscow.

To automate the experiment a multithreaded Java-based application was developed that runs on the server FREEnet and collects statistics simultaneously from two points. This fact gives us the opportunity to assess site characteristics Moscow -- Amsterdam in both directions (see Fig~\ref{f1}). Based on the experimental data, we calculated the available bandwidth of the site and investigated the dependence of measurement errors on the number of measurements in both directions.

The current system of test traffic measurement (TTM) service has different functions such as measurement of one-way packet delays, tracing routes between the test box and major DNS servers, etc. The measuring point RIPE test box which has a GPS receiver to synchronize the clock with an accuracy of about 2 microseconds is relatively expensive to install and maintain. But with the RIPE test box we can measure the available bandwidth up to the upper bound $\ensuremath{\bar{B}}=300$ {\em Mbps} with a relative error $\eta=10\%$. And, if we use the standard utility \textit{ping}, with a relative error $\eta=25\%$ and a precision of 1 millisecond $\Delta D=10^{-3}$, we could get results for the network available bandwidth up to $1.5$ {\em Mbps} \cite{s17}.

The RIPE TTM project mainly aims to assist providers in monitoring their network and planning capacity of channels. Now, when bandwidth is really high, the focus of measurement shifts from monitoring networks to obtaining accurate data about the status of the connection of the end user, from the analysis of major networks to obtain a sufficiently complete data about the Internet as a whole.

\begin{figure}[ht]
\centering
\includegraphics[height=4.6cm]{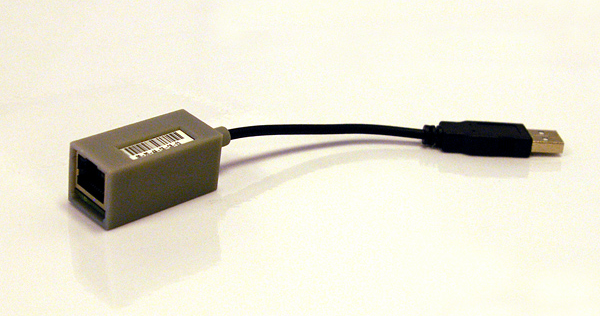}
\caption{RIPE Atlas probe}
\label{f7}
\end{figure}

To create a new measuring infrastructure project RIPE Atlas proposes to deploy a large number ($> 50000$) portable, simple and inexpensive measuring devices, known as \textit{probes} (see Fig~\ref{f7}). Each such probe is integrated into a network that is running DHCP and NAT, powered by a conventional USB-port. Probes give data about \textit{round-trip time} (RTT) and loss of sent test packets. Due to the fact that bandwidth of network channels (including channels of last mile) is steadily increasing, then to assess the QoS and creation connectivity map is extremely important to get these values in real time. The participation of SSAU in the project RIPE Atlas will allow making experiments by using a new measuring infrastructure

\section{Processing of experimental results}
After making the scheme of the experiment (see Fig~\ref{f1}) and development of appropriate software, we made a series of measurements and calculated the value of available bandwidth for both directions.

By default, the test packet size is 100 bytes, but there are special settings that allow adding testing of any RIPE test box by packets of up to 1500 bytes with desired frequency. In our case it is reasonable to add a packet size of 1100 bytes. The request on inclusion of such a mode during the experiment was approved by RIPE.

In order to gain access to the test results it is necessary to apply for remote access (\textit{telnet}) to the RIPE test box on port 9142. The data includes information about the desired delay packets of different sizes. In order to extract the data it is necessary to identify the packet on the receiving and transmitting sides. In paper \cite{s17} it was shown how to obtain the values of packet delays, knowing their numbers in the sequence.

\begin{figure}[ht]
\centering
\includegraphics[height=5.2cm]{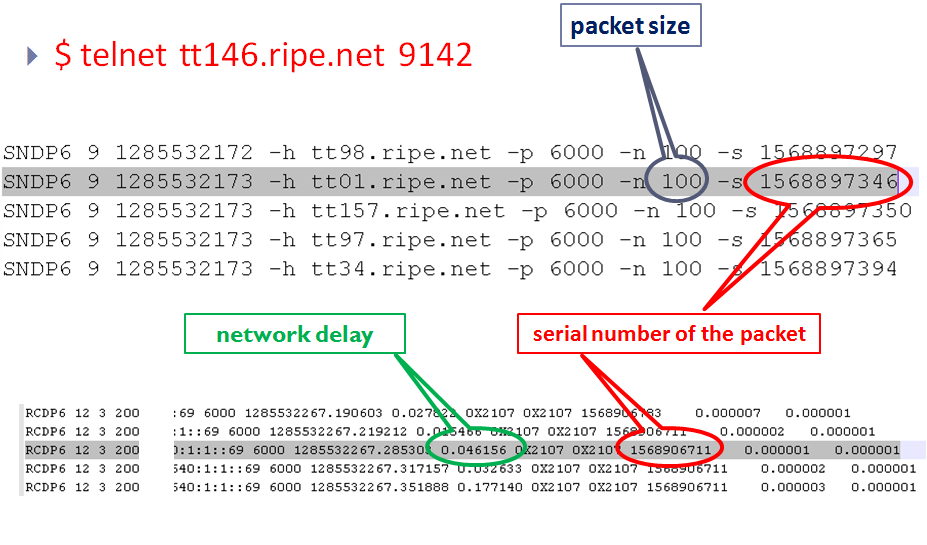}
\caption{Format of the experimental data from RIPE test box }
\label{f2}
\end{figure}

The typical format of the experimental data is shown in Fig~\ref{f2}.

Note that the model involves calculating with an average value $W_2-W_1$, so it is necessary to average several values, going consistently. In the present experiment, the averaged difference $D_{av}(1100)-D_{av}(100)$ amounted to 0.000292 seconds in the direction $tt01\rightarrow tt146$. Then the available bandwidth is:

\[B_{av}(tt01\rightarrow tt146)=\frac{8\times 1000}{0.000292}=27.4 \textit{ Mbps}\]
The average difference in the direction $tt146\rightarrow tt01$ was 0.000288 seconds. Then the available bandwidth will be:

\[B_{av}(tt146\rightarrow tt01)=\frac{8\times 1000}{0.000288}=27.8 \textit{ Mbps}\]

\begin{figure}[ht]
\centering
\includegraphics[height=5cm]{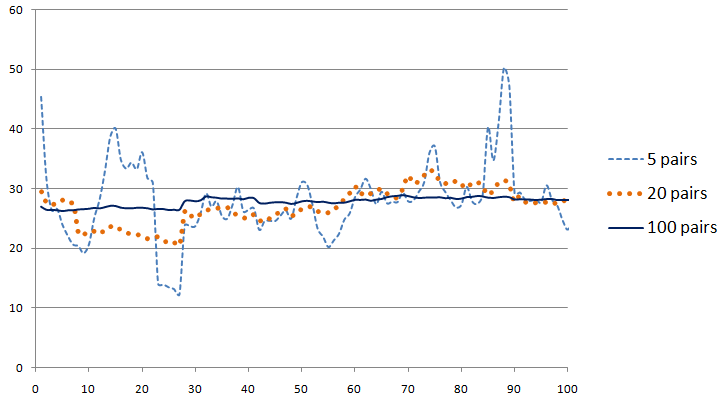}
\caption{Dependence of available bandwidth on the number of measurements in direction $tt01\rightarrow tt146$}
\label{f3}
\end{figure}

\begin{figure}[ht]
\centering
\includegraphics[height=5cm]{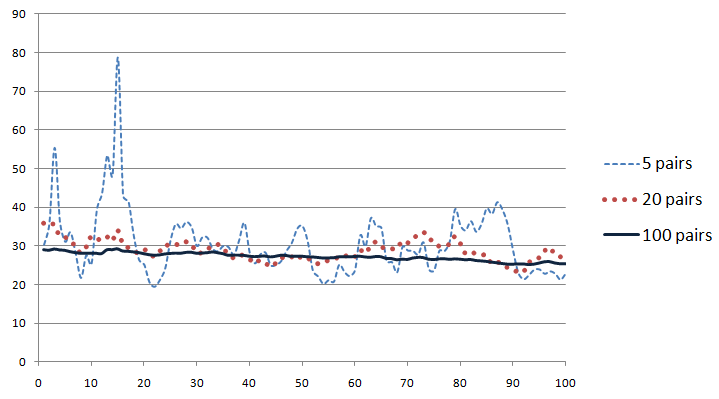}
\caption{Dependence of available bandwidth on the number of measurements in direction $tt146\rightarrow tt01$}
\label{f4}
\end{figure}

On Fig~\ref{f3} and~\ref{f4} variation of the available bandwidth, calculated for various conditions of averaging, is represented for both directions. The interval between test packets is 30 seconds, i.e. two packets come every minute.

Apparently, beatings of the calculated available bandwidth remain critical at 20 averaged values, at 50 it is less noticeable, and at 100 the values in the curve are almost equalized. There is a clear correlation between the number of measurements and the variation of the calculated available bandwidth. The same situation occured during the measurements in IPv4. The beats are caused by the variable part of the delay, which is reduced as the number of measurements increases. 

Based on data obtained from boxes tt01 and tt146 we computed the standard deviations (SD) $\sigma_n(B)$ of the available bandwidth (see Tables~\ref{t1} and~\ref{t2}).

\begin{table*}
	\centering
		\begin{tabular}{|c|c|c|c|c|c|c|c|c|c|c|c|c|}
		\hline
Number of measurements, $n$	& 5	& 10	& 20	& 30	& 40 & 50 & 60 & 70 & 80 & 90 & 100 & 200\\
		\hline
Standard deviation, & \multirow{2}*{49.3}	& \multirow{2}*{34.7} &	\multirow{2}*{24.3}	& \multirow{2}*{19.8} &	\multirow{2}*{18.3}	& \multirow{2}*{16.0} &	\multirow{2}*{14.4} &	\multirow{2}*{13.1}	& \multirow{2}*{12.1}	& \multirow{2}*{11.2} & \multirow{2}*{10.5} & \multirow{2}*{7.2}
\\ $\sigma_n(B)$ (Mbps) & & & & & & & & & & & & \\	
		\hline
The average value & \multicolumn{12}{|c|}{}
\\ of available bandwidth, & \multicolumn{12}{|c|}{27.4}
\\ $B_{av}$ (Mbps) & \multicolumn{12}{|c|}{}\\
		\hline
		\end{tabular}
	\caption{Dependence of SD on the number of measurements ($tt01\rightarrow tt146$)}
	\label{t1}
\end{table*}

\begin{figure*}
\centering
\includegraphics[height=7cm]{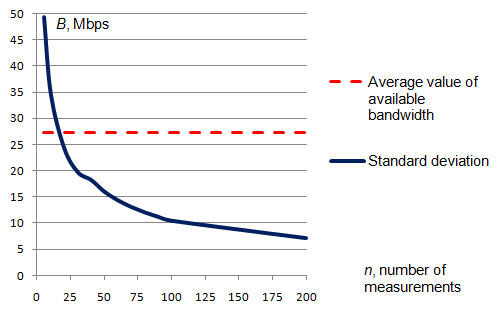}
\caption{Dependence of SD on the number of measurements ($tt01\rightarrow tt146$)}
\label{f5}
\end{figure*}

Fig~\ref{f5} shows that it is necessary to take at least 70 measurements (the delay difference for 70 pairs of packets). In this case, the calculated value exceeds twice the capacity of SD, i.e., $B\geq 2\sigma_n(B)$.

\begin{table*}
	\centering
		\begin{tabular}{|c|c|c|c|c|c|c|c|c|c|c|c|c|}
		\hline
Number of measurements, $n$	& 5	& 10	& 20	& 30	& 40 & 50 & 60 & 70 & 80 & 90 & 100 & 200\\
		\hline
Standard deviation, & \multirow{2}*{7.5}	& \multirow{2}*{5.3} &	\multirow{2}*{3.7}	& \multirow{2}*{3.1} &	\multirow{2}*{2.7}	& \multirow{2}*{2.5} &	\multirow{2}*{2.4} &	\multirow{2}*{2.2}	& \multirow{2}*{2.1}	& \multirow{2}*{2.0} & \multirow{2}*{1.9} & \multirow{2}*{1.3}
\\ $\sigma_n(B)$ (Mbps) & & & & & & & & & & & & \\	
		\hline
The average value & \multicolumn{12}{|c|}{}
\\ of available bandwidth, & \multicolumn{12}{|c|}{27.8}
\\ $B_{av}$ (Mbps) & \multicolumn{12}{|c|}{}\\
		\hline
		\end{tabular}
	\caption{Dependence of SD on the number of measurements ($tt146\rightarrow tt01$)}
	\label{t2}
\end{table*}

\begin{figure*}
\centering
\includegraphics[height=7cm]{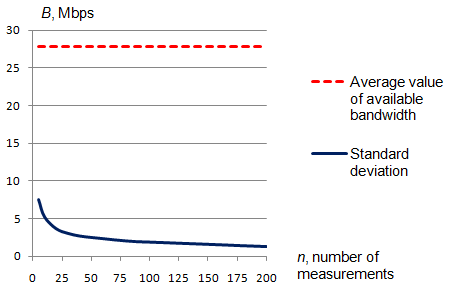}
\caption{Dependence of SD on the number of measurements ($tt146\rightarrow tt01$)}
\label{f6}
\end{figure*}

Fig~\ref{f6} shows that the variation of packet delays in the direction Moscow $\rightarrow$ Amsterdam is much smaller than in direction Amsterdam $\rightarrow$ Moscow. This could be due to the fact that the channel Moscow $\rightarrow$ Amsterdam to IPv6 is used less intensively than the channel in the opposite direction. This situation demonstrates the possibility of using the proposed model with the measurement infrastructure RIPE test box for the study of asymmetric effects in the network channels.

\section{The optimal number of measurements}
To establish the relationship between the number of measurements and permissible error a computer simulation can be carried out using a generalized generating function for describing the packet delays. In \cite{s3} it was found that the distribution of the IPv4 packet delay obeys an exponential law, for IPv6 packets such verification has not taken place yet. It was shown that the evaluation of experimental measurement errors needs to know tabulated values of relative errors in measuring the available bandwidth $\eta^{T}_{n}$ and the values of correction factors.

Correction factors can easily calculate the required number of measurements to estimate the experimental error:

 	\begin{equation}
  \eta^{exp}_{n}=\frac{\eta^{T}_{n}}{k(D_2-D_1)\cdot k(\lambda)},
  \label{e2}
	\end{equation}
where $k(\lambda)=\lambda^{exp}/\lambda^T$, $k(D_2-D_1)=(D^{exp}_{2}-D^{exp}_{1})/(D^{T}_{2}-D^{T}_{1})$ and $\eta^{T}_{n}$ come from Table~\ref{t3}, which was taken from \cite{s17}.

\begin{table}[!h]
	\centering
		\begin{tabular}{|c|c|c|c|c|c|c|c|}
		\hline
Number	& \multirow{2}*{5}	& \multirow{2}*{10}	& \multirow{2}*{20}	& \multirow{2}*{30}	& \multirow{2}*{50} & \multirow{2}*{100} & \multirow{2}*{200}
\\ of measurements, $n$ & & & & & & & \\
		\hline
Measurement &	\multirow{2}*{82.6} &	\multirow{2}*{61.1} &	\multirow{2}*{44.2} &	\multirow{2}*{35.5} &	\multirow{2}*{24.4} &	\multirow{2}*{13.9} &	\multirow{2}*{9.4}
\\ error, $\eta^{T}_{n}$ (\%)& & & & & & & \\
		\hline
		\end{tabular}
	\caption{Dependence of error on the number of measurements (IPv4)}
	\label{t3}
\end{table}

Let $\lambda^T=1000 s^{-1}$, $D^{T}_{2}-D^{T}_{1}=8\cdot 10^{-4}${\em s} \cite{s3}. The calculation will hold for the following values: $\lambda^T\approx 1000 s^{-1}$, which corresponds to $D_{av}-D_{min}=3\cdot 10^{-4}${\em s}, $D^{exp}_{2}-D^{exp}_{1}\approx 4\cdot 10^{-4}${\em s}. Substituting into Eq~\ref{e2} the values of the coefficients $k(D_2-D_1)$, $k(\lambda)$ and the desired accuracy of measurements $\eta^{T}_{n}$ we shall find the number of measurements \textit{n} required to achieve a given error.

\begin{table}[!h]
	\centering
		\begin{tabular}{|c|c|c|c|c|c|c|c|}
		\hline
Number	& \multirow{2}*{5}	& \multirow{2}*{10}	& \multirow{2}*{20}	& \multirow{2}*{30}	& \multirow{2}*{50} & \multirow{2}*{100} & \multirow{2}*{200}
\\ of measurements, $n$ & & & & & & & \\
		\hline
Measurement &	\multirow{2}*{54.0} &	\multirow{2}*{40.0} &	\multirow{2}*{28.9} &	\multirow{2}*{23.2} &	\multirow{2}*{16.0} &	\multirow{2}*{9.1} &	\multirow{2}*{6.1}
\\ error, $\eta^{T}_{n}$ (\%)& & & & & & & \\
		\hline
		\end{tabular}
	\caption{Dependence of error on the number of measurements (IPv6)}
	\label{t4}
\end{table}

During the real experimental measured quantities $D^{exp}_{2}-D^{exp}_{1}$, $\lambda^{exp}$ and $B^{exp}$ take arbitrary values, but but by using correction factors, we calculated the required number of measurements (to estimate the experimental error) on the basis of the tabulated values.

\section{Conclusion}
In this work we performed experimental checking of the model for estimating available bandwidth in IPv6 networks using a metering infrastructure RIPE test box, ensuring precision accuracy. It is shown that to achieve a minimum level of error it is necessary to take a sufficiently large number of measurements. The data obtained suggest the similarity of patterns which we observed during the measurements in the IPv4 \cite{s17}.

In the future we are going to compare the results with data measured by other methods. Unfortunately, there is a lack of utilities, which could measure the available bandwidth of different channels in IPv6 networks with acceptable accuracy \cite{s7}. It is worth noting that the organization of the experiments in a real IPv6 network is slow because of low rates of overall introduction of the new protocol. However, IPv6 is worth more and more attention from both service providers and research and education communities \cite{s2}.

\section*{Acknowledgment}
We would like to express special thanks to Dmitry Sidelnikov from the Institute of Organic Chemistry, Russian Academy of Sciences, Sean McAvoy from RIPE NCC and Mikhail Strizhov from Colorado State University for their invaluable assistance in carrying out the measurements. We also thank Andrei Sukhov, Ph.D., professor of SSAU for assistance in managing and conducting research on the subject.

\end{document}